\documentclass[aps,prb,twocolumn,superscriptaddress,showpacs,amsmath,amsfonts]{revtex4-1}
\usepackage[pdftex]{graphicx}

\begin{document}

% Use the 'preprintnumbers' class option to override journal d$E_F$aults
%\preprint{}

\title{
Topological semimetal phases manifested in transition metal dichalcogenides intercalated with 3$d$ metals
}


\author{Takeshi Inoshita}
\thanks{These two authors contributed equally.}
%\email[]{inoshita@mces.titech.ac.jp}
\affiliation{TIES, Tokyo Institute of Technology, Nagatsuta, Kanagawa 226-8503, Japan}
\affiliation{National Institute for Materials Science, Tsukuba, Ibaraki 305-0044, Japan}
\author{Motoaki Hirayama}
\thanks{These two authors contributed equally.}
\affiliation{RIKEN Center for Emergent Matter Science, Wako, Saitama 351-0198, Japan}
\author{Noriaki Hamada}
\affiliation{Faculty of Science and Technology, Tokyo University of Science, Noda, Chiba 278-8510, Japan}
\affiliation{Center for Spintronics Research Network, Osaka University, Toyonaka, 560-0043, Japan}
\author{Hideo Hosono}
\affiliation{TIES, Tokyo Institute of Technology, Nagatsuta, Kanagawa 226-8503, Japan}
\affiliation{Materials and Structures Laboratory, Tokyo Institute of Technology, Nagatsuta, Kanagawa 226-8503, Japan}
\author{Shuichi Murakami}
\affiliation{Department of Physics, Tokyo Institute of Technology, Tokyo 152-8551, Japan}
\affiliation{TIES, Tokyo Institute of Technology, Nagatsuta, Kanagawa 226-8503, Japan}

\date{\today}

\begin{abstract}
In the search for stable topological semimetals with clean band profiles, we have screened all the 3$d$ metal-intercalated transition metal dichalcogenides (3dI-TMDCs) by performing hybrid-functional-based {\it ab initio} calculations. Two classes of topological materials featuring 12 Weyl points (WPs) in the $k_z=0$ plane (without spin-orbit interactions) are identified: (a) ferromagnetic Weyl semimetals VT$_3$X$_6$ and (b) spinless Weyl semimetals MnT$_3$X$_6$ (nonmagnetic), where T=Nb, Ta; X=S, Se.  VNb$_3$S$_6$, prototypical of class (a), is half-metallic with only two bands crossing at the Fermi level to form WPs.  If spin-orbit interactions are included, VNb$_3$S$_6$ is a Weyl semimetal with {\it only two WPs} for magnetization $\mathbf{m}$ in the basal plane, whereas it is a ferromagnetic topological insulator for $\mathbf{m}$ normal to the plane.  MnNb$_3$S$_6$ in the nonmagnetic phase is essentially a spinless version of VNb$_3$S$_6$ featuring an equally clean band profile.  Interestingly, both materials exhibit a quasinodal line connecting the WPs, which we explain in terms of the quasimirror symmetry and the orbital nature of the bands.  3dI-TMDCs are chemically and thermally stable stoichiometric compounds containing no toxic elements and are a viable platform for the study of topological condensed-matter physics.
\end{abstract}   

%\pacs{81.05.Zx,73.21.-b,75.50.-y,31.15.A-}
%  Subject Area List
%    Condensed Matter Physics
%    Materials Science
%    Interdisciplinary Physics
%    Computational Physics
%\keywords{}

\maketitle

Introduction: The topological view of condensed matter, shaped gradually in the latter half of the 20th century, came to the forefront of condensed-matter research with the discovery of topological insulators.\cite{HK,QZ}  The field has since been evolving rapidly with target materials expanding from insulators to semimetals.\cite{SM,RevWeyl1, RevWeyl2,RevWeylDirac,RevTSM,RevNLSM1,RevNLSM2}   Topological semimetals can be broadly classified into Dirac semimetals (DSMs), Weyl semimetals (WSMs), and nodal-line semimetals (NLSMs). 

DSMs are characterized by having two energy bands, each spin-degenerate, crossing linearly near the Fermi level $E_F$ at isolated points in the Brillouin zone (BZ).  The crossing point (Dirac point) is therefore fourfold degenerate, which is possible only if the system respects both time-reversal and spatial-inversion symmetries.  If one or both of these symmetries is broken, a Dirac point splits into two doubly degenerate points, turning the material  into a WSM.  The degeneracy points (Weyl points or WPs) have chirality $C=\pm1$ depending on whether the node is a sink or source of Berry curvature.   The total chirality of the BZ must vanish; therefore, WPs come in pairs with opposite signs of $C$.  The degeneracy is lifted only when WPs of opposite chiralities approach each other and annihilate in pairs.  Therefore, WPs are robust against weak perturbations.  In contrast to DSMs and WSMs having degeneracies at discrete $k$ points, NLSMs have degeneracy along a one-dimensional path in the BZ.  Such a line degeneracy arises, for example, if the crystal has a mirror plane on which the bands have different mirror eigenvalues.

A salient feature of topological materials is the emergence of surface states associated with their bulk topology.\cite{RevWeylDirac}  DSMs and WSMs have topologically protected surface states called Fermi arcs, which give rise to quantum oscillations in magnetotransport or quantum interference in tunnel conductance.   NLSMs tend to have rather flat surface bands in the shape of drumheads inside or outside the nodal loop projected onto the surface BZ.   High-temperature superconductivity and special collective modes involving drumhead surface states have been predicted.\cite{Heikkila, SpecCollMode}

The material realization of topological semimetals has been spearheaded by {\it ab initio} calculations.   Density functional theory calculations have successfully identified inversion-breaking WSMs, such as transition metal monopnictides\cite{TaAs_1,TaAs_2} and trigonal Te and Se \cite{HirayamaTe}, and time-reversal-breaking WSMs such as Co$_2$TiX (X=Si, Ge, Sn)\cite{C2X_1} and Co$_2$ZrSn.\cite{ZrCoSn}  As regards DSMs, Na$_3$Bi\cite{Na3Bi} and Cd$_3$As$_2$ \cite{Cd3As2} were {\it discovered} by calculations.   The NLSMs identified include Ca$_2$As (and related compounds),\cite{Ca2As} fcc alkaline earth metals,\cite{HirayamaCa} and  Y$_2$C.\cite{Y2C1,Y2C2}  (Extensive lists of predicted materials are given in recent reviews.\cite{RevWeylDirac,RevTSM, RevNLSM2}) Such materials, however, are often unstable under ambient conditions or may require the tuning of $E_F$ to the band-crossing points by pressure or alloying.   In time-reversal-breaking materials, magnetic domain formation can be a hindrance.  Also, materials with only two bands crossing at $E_F$ are preferable since the presence of intervening non-topological bands would mask topological effects.  Further effort is needed to discover topological materials with more desirable electronic structures.

In this Rapid Communication, we revisit 3$d$ metal-intercalated transition metal dichalcogenides (3dI-TMDCs), studied intensively from the 1970's to the 1980's, and show that V/Mn intercalation compounds are topological semimetals with fascinating features. 3dI-TMDCs are interesting in that  (i) they form stoichiometric compounds that are chemically and thermally stable, (ii) their structures are well understood, (iii) most of them show a nonmagnetic-magnetic transition at temperatures below $\sim$100 K, which allows the realization of both time-reversal-symmetric and -asymmetric states, (iv) the carrier density and magnetism can be tuned readily by alloying or the selection of intercalants, and  (v) a narrow intercalant (3$d$) band, in which $E_F$ is located, overlaps a hybridization-induced energy gap within the 4$d$ manifold of the host. The last point (v) is particularly relevant, indicating the possibility that the 3$d$ and 4$d$ bands cross near $E_F$ with no other band intervening in the same energy range.

We have searched for topological band crossings near $E_F$ in all the 3dI-TMDCs using hybrid-functional-based {\it ab initio} calculations.   The search revealed that VT$_3$X$_6$ in the ferromagnetic phase ($T<T_C$, Curie temperature) and MnT$_3$X$_6$ in the nonmagnetic phase  ($T>T_C$)  (T$=$Nb, Ta; X=S, Se) are WSMs with a quasinodal line on the $k_z=0$ plane [without a spin-orbit interaction (SOI)], although they lack inversion or mirror symmetry.  We attribute these quasinodal lines to a quasimirror symmetry, induced by in-plane twofold rotation axes and the orbital characters of the crossing bands.  We focus our discussion on VNb$_3$S$_6$ and MnNb$_3$S$_6$.  [The band structures of all the screened materials are shown in the Supplemental Material (SM). \cite{SupplM}]

\begin{table*}
\caption{\label{T1}Experimental magnetic properties of MNb$_3$S$_6$ (M: 3$d$ intercalants from Ti to Ni) compiled in the review by Friend and Yoffe. \cite{FriendYoffe} Here NM, FM, and AFM denote nonmagnetic, ferromagnetic and antiferromagnetic, $T_C$ is the transition temperature, and $N_d$ and valence are the number of 3$d$ electrons and the charge state of M, respectively.}
% ref.~\onlinecite{FriendYoffe}.}
%\begin{ruledtabular}
\begin{ruledtabular}
 \begin{tabular}{lccccccc}
 & Ti & V & Cr & Mn & Fe & Co & Ni \\
\colrule
magnetic order (easy axis) & NM & FM (ab) & FM (ab) & FM (ab)& AFM (c) & AFM (-)& AFM (-) \\
$T_c$  (K) & - & 55 & 115 & 40 & 45 & 25 & 90  \\
$N_d$ (valence) & 1 (+3) & 2 (+3) & 3 (+3) & 5 (+2)& 6 (+2)& 7 (+2)& 8 (+2)\\
\end{tabular}
\end{ruledtabular}
%\end{ruledtabular} 
%\footnotetext[1]{ Ref.~\onlinecite{FriendYoffe}}
 \end{table*}

TMDCs and their 3$d$ intercalation compounds: Layer-structured TMDCs, TX$_2$ (T=transition metal, X=chalcogen), are hexagonal crystals consisting of X-T-X sandwiches [`host' in Fig. 1(a)] stacked in the $c$ direction.  While the intrasandwich bonding is covalent and strong, the intersandwich bonding is weak. \cite{Kolobov}   This anisotropy allows the electronic structures of TMDCs to be described by the monolayer model.    Figure 1(b) shows the density of states of a NbS$_2$ monolayer.  Near $E_F$, the dispersive 4$d$ bands of Nb overlap each other, but the lowest band with the $d_{z^2}$ character hybridizes with the higher-lying 4$d$ bands to create a gap.   Because this split-off 4$d_{z^2}$ band is half-filled, the system is a metal.  Among the many TMDCs, four (NbS$_2$, NbSe$_2$, TaS$_2$, TaSe$_2$) can be intercalated with 3$d$ transition metals (Ti, V, Cr, Mn, Fe, Co, Ni) to form stoichiometric compounds MT$_3$X$_6$.\cite{FriendYoffe}   

%Fig_1
\begin{figure}
\includegraphics[width=8 cm]{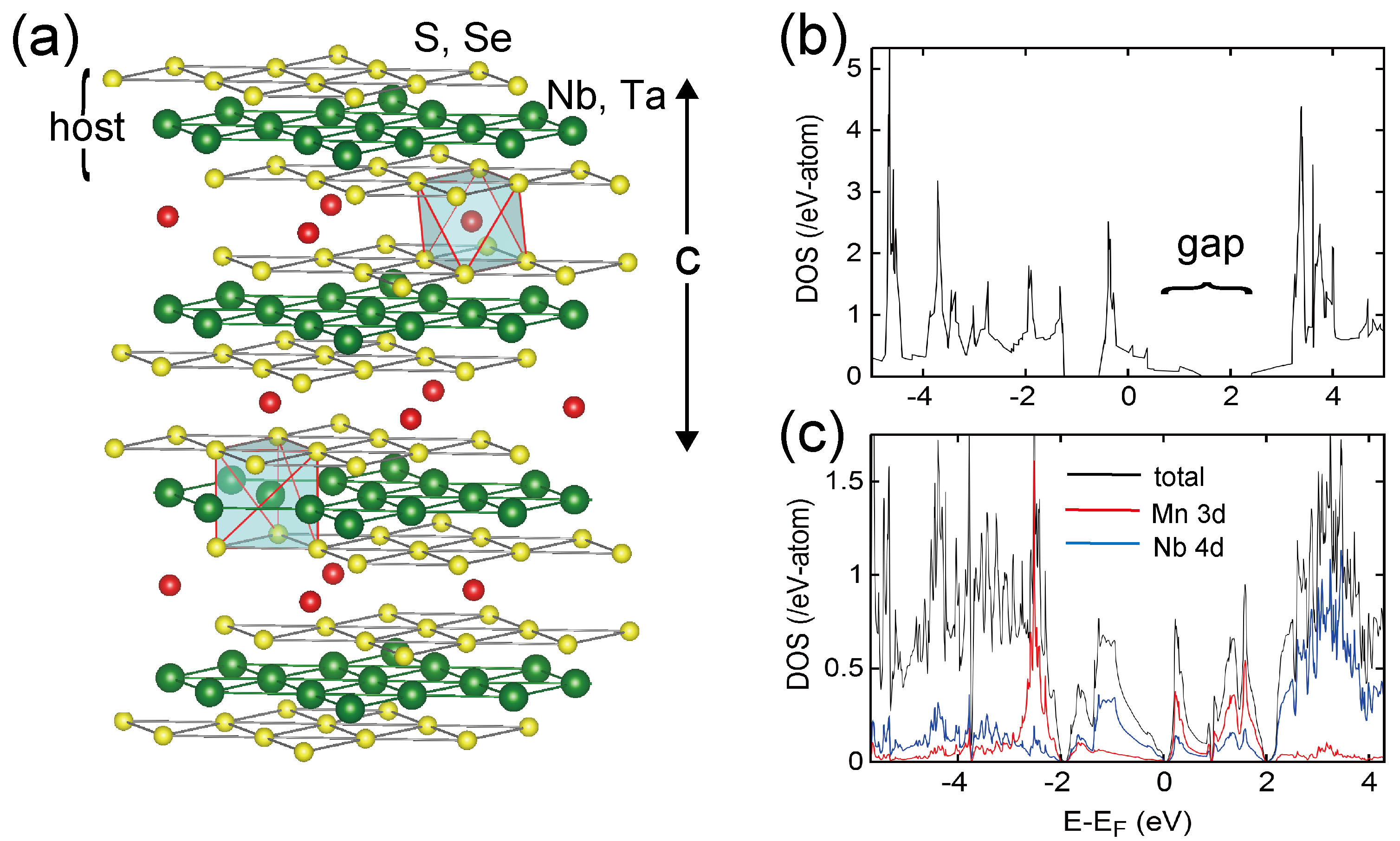}
\caption{\label{Fig1}(a) Crystal structure of TMDCs intercalated with 3$d$ transition metals MT$_3$X$_6$.  The green and yellow spheres denote the Nb/Ta and S/Se atoms, respectively, in the host layers, and the red spheres represent the intercalants.  A unit cell contains two host sandwiches (TX$_2$) and two intercalant (M) layers.  (b) Density of states calculated for a NbS$_2$ monolayer.  (c) Density of states calculated for MnNb$_3$S$_6$ in the nonmagnetic phase.  The horizontal axes for the two plots are offset to align the host energy bands in the two figures.}
\end{figure}

All the MT$_3$X$_6$ compounds crystallize in the same structure without inversion or mirror symmetry (space group P6$_3$22, No. 182) shown in Fig.~1(a). Of the symmetry elements of P6$_3$22, essential for our later discussion are the two classes of twofold rotations around the in-layer axes, $C_2''$ (axes parallel to $<$100$>$) and $C_2'$ (axes parallel to $<$120$>$).\cite{Dres}  Within a sandwich, the Nb/Ta ions take trigonal prismatic coordination with the chalcogens.   The adjacent sandwiches are stacked so that the Nb/Ta ions sit on top of each other, whereas the chalcogen layers separated by an intercalant layer are rotated by 60$^{\circ}$ with respect to each other, causing the intercalants to take octahedral coordination with the chalcogens.   The unit cell comprises two TX$_2$ sandwiches and two intercalant layers; the in-layer unit cell is $\sqrt 3 \times \sqrt 3$ relative to the host unit cell.  As the band structures of the hosts are similar, the electronic structures and properties of 3dI-TMDCs are determined mainly by the intercalant species.  These properties, particularly the magnetic properties, attracted considerable attention from the 1970's to the 1980's.\cite{FriendYoffe}   Table I summarizes the key experimental results for the NbS$_2$-derived compounds. 

Figure 1(c) shows the density of states of MnNb$_3$S$_6$ in the nonmagnetic phase ($T>T_c$).  Whereas the host 4$d$ bands are largely unaffected by intercalation, the Mn 3$d$ orbitals form rather narrow bands near $E_F$, spanning the energy gap of the host.   Two of the Mn electrons transfer to the host 4$d_{z^2}$ band and  stabilize the system.   This rigid band picture generally holds true for all members of the 3dI-TMDC family.  The number of 3$d$ electrons increases from Ti to Ni (Table I), giving rise to material-specific magnetic order.  The compounds undergo a nonmagnetic-magnetic transition at temperatures between 25 and 115 K.  Similar magnetic trends have also been observed in 3dI-TMDCs derived from other hosts.\cite{SupplM}

Calculation method: We calculated the electronic structures of all the 3dI-TMDCs with known structures using the hybrid-functional (HSE06) method \cite{HSE,HSE06,Mars, PBE} in the projector augmented wave (PAW) framework, as implemented in the Vienna {\it ab initio} simulation package (VASP).\cite{Kr,Kr2,Bl,Kr3}  [See SM for calculational details and a comparison between the hybrid-functional and generalized gradient approximation (GGA)+$U$ results.\cite{SupplM}] The experimental crystal structures\cite{ParkinFriend, FriendBielYoffe,TiCr, VMn, FeNi,Co} were assumed. 

%Fig_2
\begin{figure}
\includegraphics[width=8 cm]{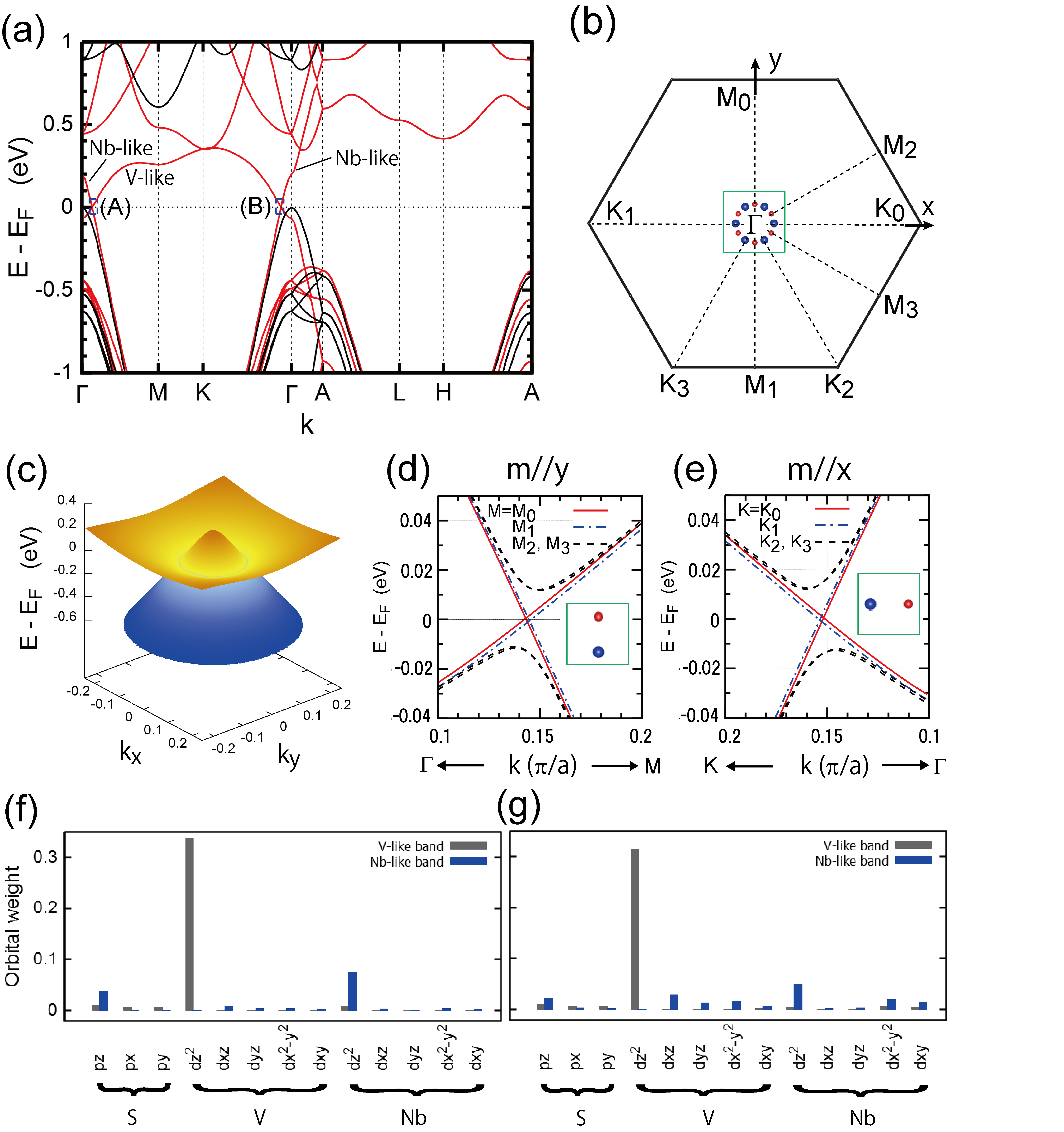}
\caption{\label{Fig2}(a) Band structure of VNb$_3$S$_6$ in the ferromagnetic phase without SOI.   The majority-spin (minority-spin) bands are shown in red (black). 
(b) Position of the WPs in the absence of SOI in the BZ ($k_z$=0 plane).  The small red (large blue) spheres denote WPs with $C=+1$ ($-1$). (c) Two-dimensional band plot in the $k_z$=0 plane in the absence of SOI.  Note that the two bands are highly isotropic and appear to touch each other along a circular loop. (d) Band structure including SOI corresponding to box (A) in (a) for nonequivalent $\Gamma$-M paths [$\Gamma$-M$_0$, $\cdots$, $\Gamma$-M$_3$ in (b)] for magnetization $\mathbf{m}$ parallel to $\mathbf{y}$ (M$_1$-$\Gamma$-M$_0$). Of the 12 WPs in the absence of SOI (b), one on $\Gamma$-M$_0$ ($C=+1$, red sphere) and another on $\Gamma$-M$_1$ ($C=-1$, blue sphere) remain. ($k$ is in units of $\pi/a$ where $a$ is the in-plane lattice constant.) (e) Band structure including SOI corresponding to box (B) in (a) for $\mathbf{m}$ parallel to $\mathbf{x}$ (K$_1$-$\Gamma$-K$_0$). Two WPs remain, one on $\Gamma$-K$_0$ ($C=+1$, red sphere) and the other on $\Gamma$-K$_1$ ($C=-1$, blue sphere).   In (d) and (e), the positions and chiralities of the Weyl points are indicated in the boxes corresponding to the box in (b).  (f)-(g) Atomic orbital weights (modulus squared of the coefficients per atom) in the wave functions at $\Gamma$ and at a WP on $\Gamma$-M, respectively.}
\end{figure}

Results and discussion: Figure~2(a) shows the band structure of VNb$_3$S$_6$ in the ferromagnetic phase without SOI.  (Since spin and spatial freedoms are completely decoupled in the absence of SOI, the rotational and screw symmetries are preserved irrespective of the direction of the magnetization.)  Electrons transferred from the intercalated V fill up a minority-spin 4$d$ band of the host (black curve with a peak slightly below $E_F$ at $\Gamma$) and bring $E_F$ to the points (WPs) where two majority-spin bands (red curves) cross.   Among the two crossing bands, the steep band with a peak at $\Gamma$ mainly has the character of Nb 4$d_{z^2}$, whereas the narrower band with a minimum at $\Gamma$ consists predominantly of V 3$d_{z^2}$; we call them ``Nb-like band'' and ``V-like band'', respectively.  Note the extremely simple and clean half-metallic band structure with only two minority-spin bands at $E_F$.   There are 12 WPs arranged circularly on the $k_z=0$ plane [Fig.~2(b)], six with $C=+1$ on $\Gamma$-M and six with $C=-1$ on $\Gamma$-K.  The crossing on $\Gamma$-M  ($\Gamma$-K) arises because the little group on the line comprises an in-plane twofold rotation $C_2'$ ($C_2''$),\cite{Dres} whose eigenvalues are different ($+1$ vs $-1$) for the two bands.  [The eigenvalues are indicated in Fig.~3(a).]  The two types of WPs differ in energy by only 0.6 meV with $E_F$ located between them. The band degeneracy is lifted off the symmetry lines, but the gap remains remarkably small ($\ll$ 0.1 meV) along a circular loop passing through the WPs [Fig.~2(c)].

%Fig_3
\begin{figure}
\includegraphics[width=8 cm]{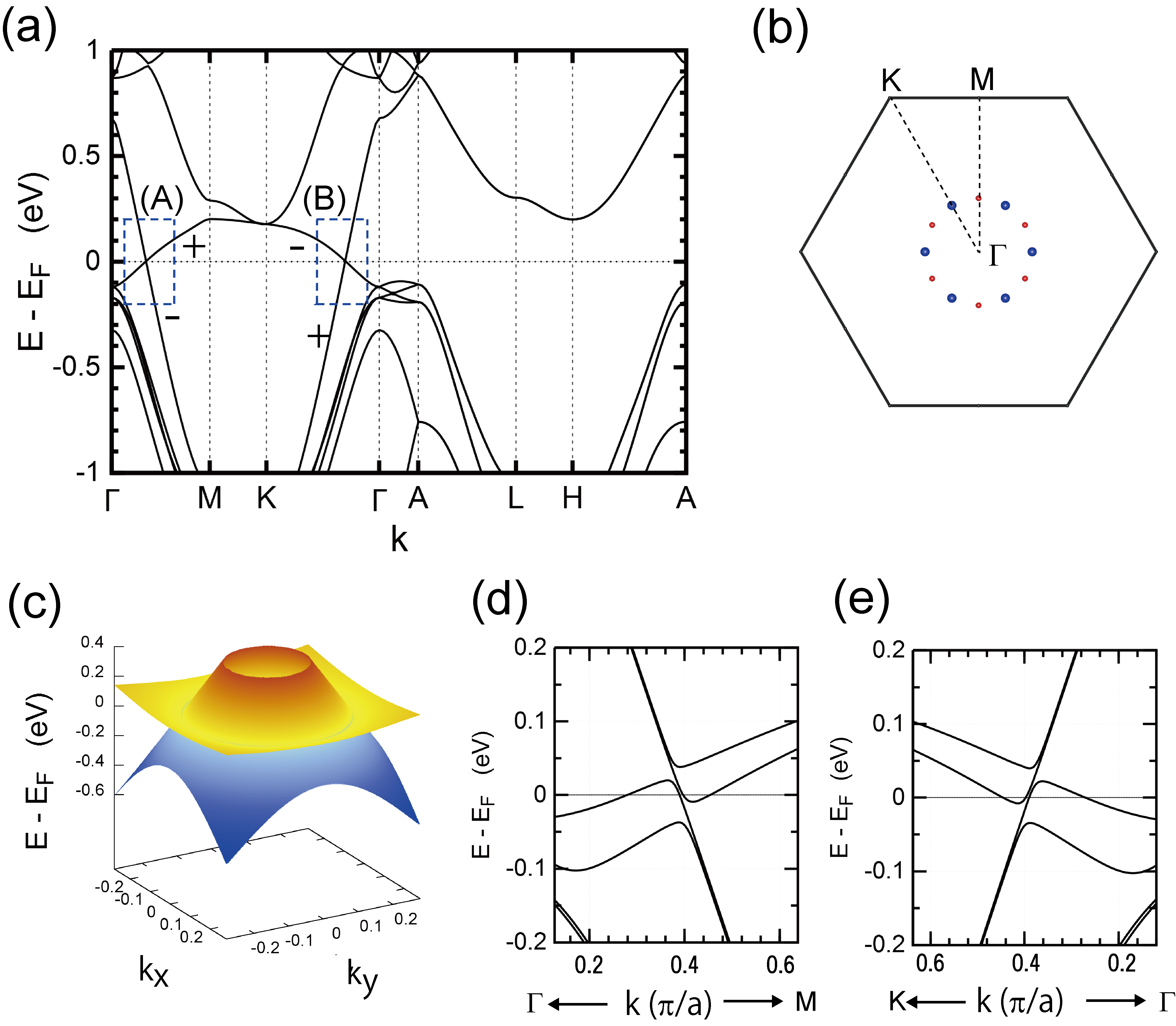}
\caption{\label{Fig3}(a) Band structure of MnNb$_3$S$_6$ in the nonmagnetic phase calculated without SOI.  The corresponding density of states is shown in Fig.~1(c).  The positive and negative signs denote the eigenvalues of $C_2'$ and $C_2''$, respectively. 
 (b) WPs in the BZ ($k_z$=0 plane) in the absence of SOI.  Each small red (large blue) sphere denotes a set of two
spinless WPs, one for each spin channel, with $C=+1$ $(-1)$.  (c) Band dispersion in the $k_z$=0 plane in the absence of SOI.  (d)-(e) Energy band structure calculated with SOI in box (A) and (B), respectively, in (a).}
\end{figure}

%Fig_4
\begin{figure}
\includegraphics[width=8 cm]{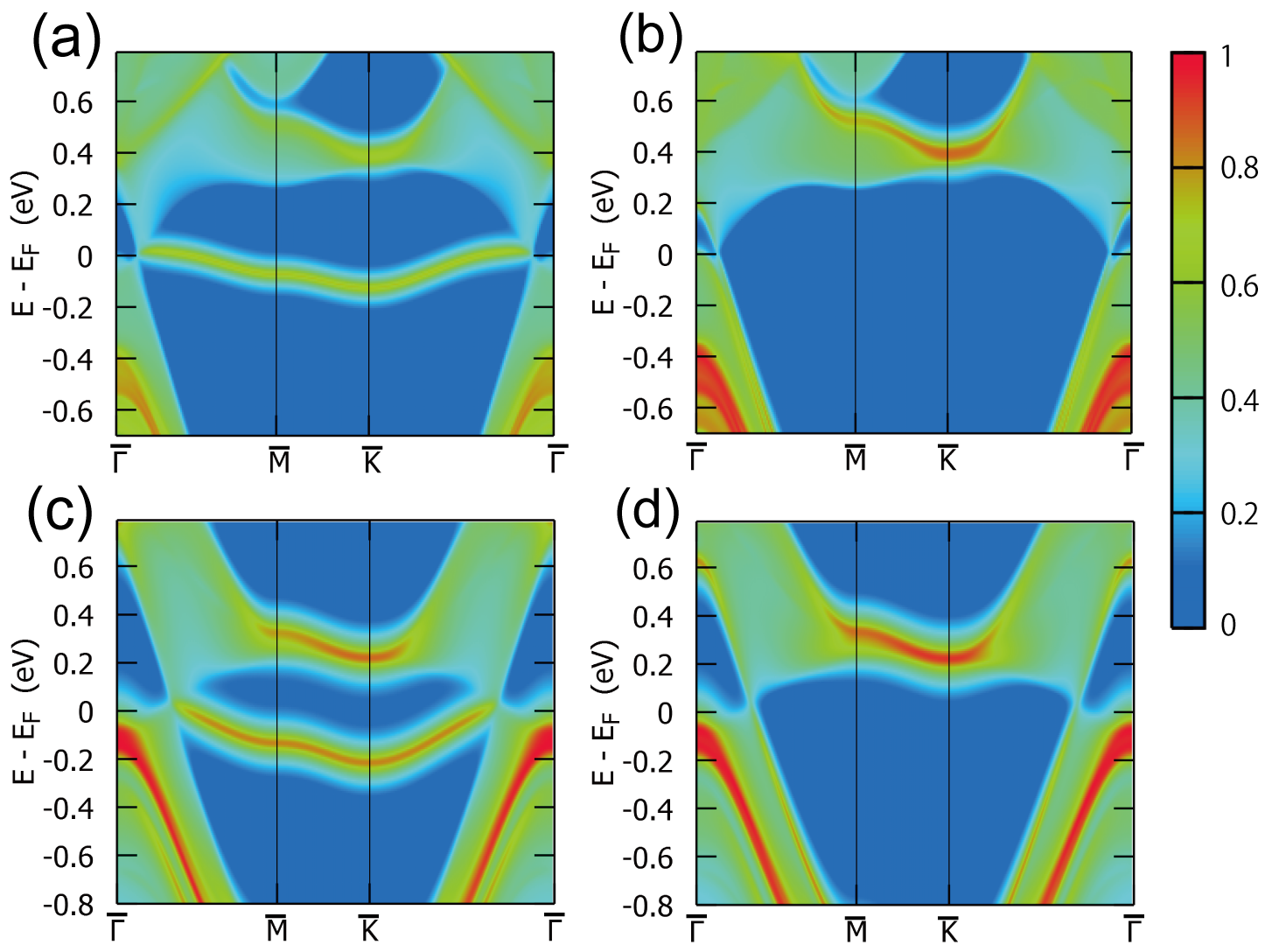}
\caption{\label{Fig4}(001) surface spectral function in the absence of SOI for the (a) V-terminated and (b) S-terminated VNb$_3$S$_6$ and the (c) Mn-terminated and (d) S-terminated  MnNb$_3$S$_6$. For VNb$_3$S$_6$ only the majority spin channel is plotted because there is no minority spin surface state near $E_F$. For V/Mn termination, surface states originating from the bulk WPs are clearly seen around $E_F$.  No such surface band is visible near $E_F$ for S-terminated surfaces.}
\end{figure}

The orbital weights of the two bands at $\Gamma$ and at a WP on $\Gamma$-M are shown in Figs.~2(f) and (g), respectively.  The plots reconfirm that on either $k$ point, the V-like band consists overwhelmingly of V $d_{z^2}$, whereas the Nb-like band is made of Nb $d_{z^2}$ and S $p_z$ orbitals.  If one ignores S and considers a hypothetical crystal made of V and Nb, it has a horizontal mirror plane passing through a V site [Fig.~1(a)].  This situation is reminiscent of  mirror-protected NLSMs whose line degeneracy results from the crossing of bands with different mirror eigenvalues on the mirror plane.  Crude as this reasoning may seem, we indeed found that the wave function of the V-like (Nb-like) band at small $k$ is nearly symmetric (antisymmetric) with respect to the V plane.  This quasimirror symmetry explains the quasinodal-line degeneracy observed in VNb$_3$S$_6$ (and similar 3dI-TMDCs).  In SM, it is shown that the quasimirror symmetry stems from the in-plane twofold rotation axes combined with the orbital nature of the bands.\cite{SupplM}

When SOI is taken into account, the band structure depends on the direction of the magnetization $\mathbf{m}$. Figure~2(d) shows the result  in box (A) in Fig. 2(a) obtained for $\mathbf{m} \parallel \Gamma$-M$_0$ [see Fig. 2(b)]. Since the system is still $C_2$-symmetric around this line, the WPs on $\Gamma$-M$_0$ [the crossing between the red solid curves in (d)] and $\Gamma$-M$_1$ (the crossing between the blue dashed-dotted curves) remain,  but the WPs on the other $\Gamma$-M lines have disappeared.  As regards the WPs on $\Gamma$-K, they are all removed by SOI owing to the loss of $C_2$ symmetry therearound.  Therefore, VNb$_3$S$_6$ in this configuration is a {\it minimal WSM hosting only two WPs} with $C=+1$ and $-1$.  A similar situation arises when $\mathbf{m}$ is aligned with $\Gamma$-K, as can be seen in Fig. 2(d) for $\mathbf{m} \parallel \Gamma$-K$_0$.  Only two WPs remain: one on $\Gamma$-K$_0$  ($C=1$) and the other on $\Gamma$-K$_1$ ($C=-1$).  When $\mathbf{m}$ is rotated in the basal plane, the WPs are found to persist and rotate around $\Gamma$ following the rotation of $\mathbf{m}$.   If, on the other hand, $\mathbf{m}$ is tilted out of the basal plane, the two WPs are annihilated, transforming the system into a ferromagnetic topological insulator with a small gap of 1 meV. (For more results including anomalous Hall conductivity, see SM.\cite{SupplM})

MnNb$_3$S$_6$ in the nonmagnetic phase is essentially a spinless version of VNb$_3$S$_6$ [Fig.~3(a)].  Time-reversal symmetry together with the neglect of SOI cause each of the two bands to be spin-degenerate, and there are spinless WPs at the crossing points.  The WPs on $\Gamma$-M and $\Gamma$-K are only 1 meV apart, and $E_F$ is located between them.  The bands are rather isotropic in the $k_z$=0 plane and nearly  degenerate along a loop, making MnNb$_3$S$_6$ a quasinodal-line semimetal similar to VNb$_3$S$_6$ [Fig.~3(c)].  With the inclusion of SOI, the spin degeneracy is lifted and the WPs disappear, but there is no global band gap [Figs.~3(d) and 3(e)].  

Figures~4(a) and 4(b) show the V-terminated and S-terminated (001) surface spectral functions, respectively, for VNb$_3$S$_6$ in the absence of SOI.  Only the minority spin channel is plotted.  In performing this calculation, we constructed tight-binding Hamiltonians $H$ from the hybrid-functional results using the WANNIER90 code,\cite{Mar, Mostofi} and from $H$ obtained the Green's functions for semi-infinite space\cite{Green} using the WannierTools package.\cite{WTools}　On the V-terminated surface [Fig. 4(a)], there is a flat surface state around $E_F$ that connects the projections of the bulk WPs onto the surface BZ.  
It has the shape of a drumhead, characteristic of NLSMs, rather than a Fermi arc usually seen in WSMs. This is consistent with the quasi-NLSM character of the bulk.  For the S-terminated surface [Fig. 4(b)], there is no clearly discernible surface state connecting the bulk WPs.  Similar results were obtained for MnNb$_3$S$_6$ [Figs.~4(c) and 4(d)]: A topological surface band is seen only in the Mn-terminated surface.

Concluding Remarks: We have screened all the 3dI-TMDCs from the topological perspective and identified (a) time-reversal-breaking WSMs VT$_3$X$_6$ ($T<T_C$) and (b) spinless WSMs MnT$_3$X$_6$ ($T> T_C$), where T=Nb, Ta  and X=S, Se.  VNb$_3$S$_6$, representative of class (a), is half-metallic, where only two bands are present at $E_F$ and cross each other.  When SOI is taken into account, VNb$_3$S$_6$ is a minimal WSM with two WPs or a ferromagnetic topological insulator, depending on whether the magnetization is in the basal plane or perpendicular to it.   This finding, combined with the soft magnetism reported in the literature,\cite{FriendYoffe} suggests the interesting possibility of tuning the topological signature of the material with a small external magnetic field. MnNb$_3$S$_6$ in the nonmagnetic phase, belonging to class (b), is essentially a spinless version of VNb$_3$S$_6$, a WSM with an equally clean and simple band profile.  The WPs in these materials are located at or within a few meV of $E_F$, implying that there is no need for Fermi level tuning.    

Magnetic 3dI-TMDCs, except Fe compounds, are soft magnetic materials\cite{FriendYoffe} with the magnetization direction and domain size readily controllable by the application of a weak magnetic field.  This makes ferromagnetic 3dI-TMDCs enticing, because magnetic domains tend to suppress topological properties.  

On a final note, clean topological band structures similar to those of VNb$_3$S$_6$ and MnNb$_3$S$_6$ are also found in compounds derived from other hosts,\cite{SupplM} making 3dI-TMDCs an attractive platform for the study of topological condensed-matter physics.

\begin{acknowledgments}
The following resources at NIMS were essential for our study:  the crystal structure database AtomWork-Adv and the automated band calculation system CompES-X (for the initial screening of materials),\cite{MatNavi} and the Numerical Materials Simulator (for numerical computation).  This work was supported by the MEXT Elements Strategy Initiative to Form Core Research Center and the JST ACCEL Program.  H. H. and N. H. acknowledge MEXT/JSPS KAKENHI Grants No. 17H06153 and No. 18K03550, respectively.   
\end{acknowledgments}

\end{document}